# DELOCALIZATION OF A DYNAMIC SYSTEM WITH PRESERVATION OR SELF-RESURECTION OF INFORMATIONAL FUNCTIONALITY

*Ghost machines or can a system survive its destruction*

Vadim Astakhov, Tamara Astakhova
UCSD astakhov@ncmir.ucsd.edu, vadim_astakhov@hotmail.com

Abstract

We analyzed the problem of a dynamic system delocalization due to changes in the system environment - *universe* and system architecture. We developed a Delocalization of Dynamic Cores model to analyze the migration of functional properties in open information and dynamic systems undergoing architecture transition and modifications. Information geometry and topological formalisms are proposed to analyze informational dynamic systems. Different physical and holographic models are proposed to construct systems able to conserve their functional properties under delocalization transition. We found several constraints for the system environment - universe which conserve the dynamic core system functionality under transition from localized explicit implementation of functions to the implicit distributed implementation.

INTRODUCTION

In the modern science and engineering, open information and dynamic systems such as various computer networks, biological neural network, gene network, social network and many others provide infrastructure for global communication and information sharing within a specified domain. These systems strongly depend on topological properties of underlying networks as well as on dynamic properties represented by hierarchy of communication pathways. These networks are not uniform sets of communicating units. They usually have hierarchical structures in which some nodes produce much more complex dynamic behavior compared to others. The situation becomes even more complicated at the *network to network* communication layer for brain neural networks, interactions among social groups or enterprise software systems where various nets run distributed applications such as distributed relational databases, grid, and J2EE/EJB and .NET platforms while others run thin client applications like browsers.

Current mathematical formalism to analyze these communication systems initially evolve with an implicit assumption that we can call "point-to-point" communication paradigm. This implies a simple use-case in which one system element sends an information stream to another one through the network. That paradigm is fine and lead to great achievements in the world of software engineering such as client-server architecture in which most applications operate as either a client or a server. The underlying global network architecture and protocol dynamics imply many assumptions for the design of such systems.

We argue that even if this paradigm applies for many applications it is not successful in some large-scale systems. We demonstrate that global properties of the system and surrounding environment might lead to conservation of the system functionality through the process that we call "*Delocalization Of Dynamic Cores*". We will demonstrate several examples of peer-to-peer communication systems as well as examples from physics and neural network dynamics.

DELOCALIZATION OF DYNAMIC CORES

With growing interest to ultra-large-scale distributed systems such as an internet, biological and social networks the global properties and architecture of the environment should be reviewed to support the software challenges of the future. That will lead to various implications for underlying network topology and design of network communication protocols.

To address those challenges, we develop a formal mathematical model that will help us analyze the complex dynamics of information processes. We extended and modify some methods from the theory of *dynamical systems* to make it applicable for analysis of ultra-large-scale systems.

A *dynamical system* is a mathematics concept in which a fixed rule describes the time dependence of a point in a geometric space. We employ the concept of "informational geometry" [1] as the geometric *space* and introduced an *informational manifold* and *system state* as a point on the manifold to describe the dynamics of information processes.

A *system state* is determined by a collection of numbers that can be measured. Small changes in the state of the system correspond to small changes in the numbers. The numbers are also the coordinates of a geometric space—a manifold. The *evolution rule* of the dynamical system is a rule that describes what and how future states follow from the current state.

The first problem we are interested in is to analyze conservation of information properties of a dynamic system and survival or self-resurrection of those properties under the transition of the underlying physical infrastructure from one medium to another.

Tornado gives us a simplest physical example of the dynamic system with changing elements but conserved functionality such as a rotor of particle velocity. The more complex examples of the function transition in computer science are migration network from a wired to optical network or from the IPv4 to IPv6 protocol. These migrations usually encapsulate the details of the underlying medium architecture and do not effect higher level applications. Such an approach usually requires partitioning of layers and development of some kind of abstraction "adapter" layer between application and the medium that will accommodate application calls to the new medium. These adapter layers might lead to inefficiency in the application and hence poor performance. Such inefficiency is usually due to assumptions about some aspects of the underlying environment that are implicit in the application design. These assumptions should be unfolded by the adapter and adapted to the new medium.

Another type of migration keeps functionality intact but requires its re-implementation. As an illustration, consider an information integration network in which some node A sends a message to node B with a request to forward the message to node C through the shortest path on the network. This case can be found in many social, biological or software systems.

One way to implement such dynamical system is to have a central node-server and a network of nodes. The server "A" that can somehow calculate the shortest paths among nodes and send messages. An arbitrary node should be able to perform a few basic operations: *get message*, *read message*, and *send message to some pre-defined neighbor*. As can be easily seen, such architecture is very dependent on the central node. The whole system is down if the central node is down.

Likely, similar functionality can be implemented effectively in a so called peer-to-peer model without a central node and in which each node can performs operations: *get message*, *read message*, and *send message to all addresses*. Such system will effectively implement the same functionality as the central-server system. Due to broadcasting "*send message to all addresses*" the message will be transmitted from A to B and then from B to C through the shortest path as well as through many other paths. We can say that *find the shortest path* function is implicitly implemented in the peer-to-peer model as opposed to explicitly implemented in the central-server model. We can call it a "*ghost function*" or "*ghost machine*". We can say that "shortest path" *resurrected* in the system that lost central node but evolved to peer-to-peer model.

This example illustrates preservation of some functionality during architecture transition as a result of abundance of the server A and modification of the communication protocol. Obviously that even if some functionality survive such transition in architecture, the other can be lost. For example, modification of the protocol that will not broadcast the message but sent it to random address will not conserve *shortest path* functionality.

*Figure 1 The left figure represents central-server architecture of communication nodes where a central general server A can sends a request to pass a message from node B to node C through the pre-computed shortest path from B to C. The same functionality can be implemented by a peer-to-peer architecture (right) where each node broadcasts the message to all neighbors.*

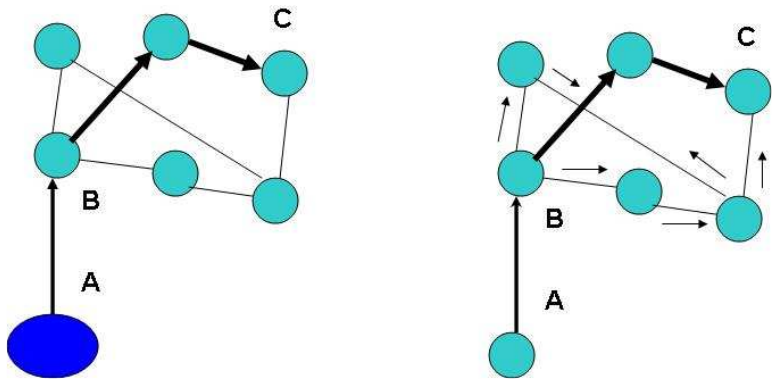

The example with "shortest path" can be extended to the information integration system - *mediator*. Usually, a *mediator* is a server that can communicate with remote data sources {D1, D2,..} and perform distributed queries. One of the key functional components of a mediator system is a *planner* that creates order-of-execution commands {Q7,Q2,Q3, ...} that should be sent to remote sources {D7, D2, D3, ...}. The planning algorithm can be quite complicated because it tries to determine which source should be queried first (D7) and what intermediate information should be send to subsequent sources {D2, D3, ...}. But such *planner* functionality also can implicitly emerge in distributed system of nodes with a few simple rules:

Starting node at A try to execute any of {Q}
Broadcast the result and all {Q} to all neighbors
A neighbor tries to execute any of {Q} and get a result, then mark message "passed me"
(if the neighbor get message that it already passed then it drop the message)
Repeat wave of broadcasts from neighbors until all {Q} have been executed
The node executed the last sub-query then returns results to A

As we can see, planning functionality will be implicitly performed as result of broadcast wave propagations. Obviously such system could consume considerable bandwidth and may be not feasible from an engineering point of view, but this example gives us a sense of the *problem of multiple implementations* for pre-defined functionality, conservation of information dynamics and emergence of functionality.

The same functionality can be implemented on different media with different dynamic properties. The question we are asking is "*what class of modifications of a dynamic information system will conserve its functional properties*" or, from another angle, "*what functional properties will survive the system transition to modified architecture.*"

We introduce some definitions:

*Dynamic Core* -- A dynamic system that consists of a set of dynamic elements causally interacting with each other and the environment in a way that lead to a high level of information integration within the system, a higher level of causal power among system elements compared to causal interactions with an environment and emergence hierarchical causal interactions within the system. The dynamic core can be seen as a functional cluster characterized by strong mutual interaction among a set of sub-groups over a period of time. It is essential that this functional cluster be highly differentiated. Examples include enterprise software agents such as EJB or COM objects, dynamic self-organization systems such as DNA and proteins, tornado super-cells, self-assembly in nano-structures such as nano-tubes and quantum dots.

*Environment* -- A set of the dynamic elements with less causal power that does not produce stable information integration. An example would be any physical medium or the whole cyber infrastructure for an informational agent.

*Delocalization* or *holographic representation/inversion* -- A process that preserve a dynamic core functionality but distribute various functional parts of a dynamic core system to characteristics of the environment.

*Universe* – term universe will be used to describe whole system including environment and dynamic cores

Dynamic core will be used to describe any dynamical system that has a sub-part acting in causal relations with each other. To measure *causal relation* some metrics considered are *information integration* and *causal power*. Dynamic core is defined as a sub-system of any physical environment that has internal *information integration* and *causal power* much higher than mutual *information integration* between the system and the environment.

Delocalization will be used for the process that leads to transition of the system in the sense of elimination of the sub-systems' *causal power* with conservation of total *information integration* for the global system. As we will demonstrate, delocalization is different from information conservation in several ways. Information can decrease that would be fine with entropy but *causal power* conservation will require that only topology of information space be conserved.

THEORY

*1. Information*

We introduce a special case of information dynamical systems or *universes* build upon two components: *core* and *environment*". Term *core* refers to a sub-system that implements some pre-defined functionality. *Environment* refers to other elements of the whole system that do not directly contribute to implementation of the functionality. The term *dynamic core* refers to the dynamic functionality that can be implemented by various sub-systems.

Generalize model [2], we consider a *information universe* - formal dynamic information system X composed of n units. Each unit can represent either distributed information objects like an element of neural, social or biological network, program object in operation memory of a computer or computer on the network. Those units can be either "on" or "off" with some probability. "On" means that this element contributes to implementation of the pre-defined function and "off" is otherwise. If these units are independent and a unit can be either "on" or "off" with equal probability then the number of states that system can occupy will be $2^n$: Every possible system state can occur with equal probability and this corresponds to a system informational entropy of $\log_2(2^n) = n$ bits. The entropy of the system $H(X)$ is simply the sum of the entropies of its individual elements $H(x_i)$ [Appendix A].

If there are any causal interactions within the system such as signals or information transfer, the number of states that the system can take will be less than the number of states that its separate elements can take. In this case, the entropy of the entire system will be less than the sum of the entropies of its individual elements.

The loss of entropy that is due to the interactions among its elements will be called the integration $I(X)$ of *universe* X.

$$I(X) = \sum H(x_i) - H(X)$$

Integration can be calculated not just for the entire system but for any of its subsets. Consider a subset j of k elements of the system X. Integration $I(X^k_j)$ measures the total statistical dependence within the subset.

As we can measure the statistical dependence within a subset, we can also measure the statistical dependence between the subset $X^k_j$ and the rest of the system $(X - X^k_j)$. The dependence can be expressed in terms of mutual information:

$$MI(X^k_j, X - X^k_j) = H(X^k_j) + H(X - X^k_j) - H(X)$$

*Figure 2. System X partitioned to subset of elements $X^k_j$ and the rest of the system $X - X^k_j$. The dashed elipse represents another possible partition.*

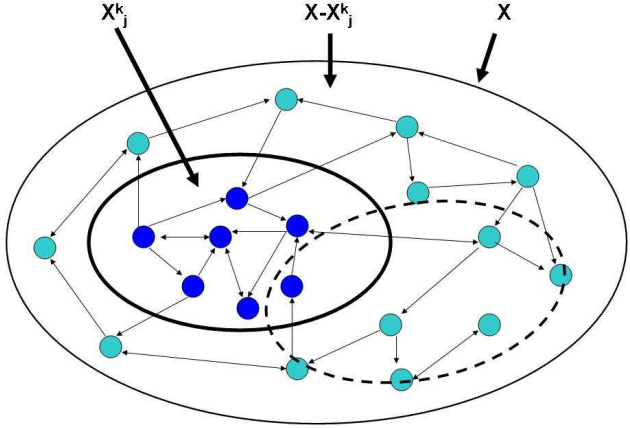

*2. Complexity and Cluster Index*

To analyze the complexity of interactions within the whole system a complexity [] function can be introduced as the sum over all subset sizes k of average mutual information between subset of size k and rest of the system:

$$C(X) = \sum_k \langle MI(X^k_j; X - X^k_j) \rangle \quad \text{the average taken over all subsets of size k}$$

The value of the average mutual information will be high if, on average, each subset can take on many different states and these states make a difference to the rest of the system.

Some subset can strongly interact within itself and much less with other regions. To analyze such a case a *cluster index* [2] was proposed to normalize and discount the subset size.

$$CI(X^k_j) = I(X^k_j) / MI(X^k_j; X - X^k_j)$$

A cluster index near 1 indicates a subset of elements that are as interactive with the rest of the system as they are within their subset. On the other hand, a cluster index much higher than 1 will indicates the presence of a *functional cluster* subset of elements that are strongly interactive among themselves but only weakly interactive with the rest of the system. A functional cluster will correspond to the subset of elements having a high cluster index that does not contain within itself any smaller subset with a higher cluster index. A system might have several functional clusters which can represent causally related information objects involved in implementation of certain functionality.

*3. Causal Power and Recursive Complexity*

The definitions for complexity and cluster index seem very generic and as such can be applicable to many statistical systems with multiple elements but at the same time they do not reflect information about causal interactions. To overcome these obstacles, we employ the concepts of *effective information* and *information integration* [2]. Effective information $EI(X^k_j \rightarrow X - X^k_j)$ between node $X^k_j$ and $X - X^k_j$ can be defined as an amount of informational entropy that $X - X^k_j$ shares with $X^k_j$ due to causal effects of $X^k_j$ on $X - X^k_j$. The simplest way to calculate $EI(X^k_j \to X - X^k_j)$ is to substitute the $X^k_j$ object by random $X^k_j(random)$ that generates random tokens then calculates mutual information between $X^k_j(random)$ and $X - X^k_j$: $EI(A \to X - X^k_j) = MI(X^k_j(random), X - X^k_j)$. Thus, if causal connections between $X^k_j$ and $X - X^k_j$ are rich and specialized then different tokens from $X^k_j$ will produce different patterns from $X - X^k_j$ and effective information will be high. Conversely if different outputs from $X^k_j(random)$ will produce little variations on $X - X^k_j$ then mutual information will be low or zero. This implies a low amount of causal interactions from A to $X - X^k_j$.

The total effective information for any bi-partition $X^k_j$ and $X - X^k_j$ can be found as a sum of the effective information for both directions:

$$EI(X^k_j \leftrightarrow X - X^k_j) = EI(X^k_j \to X - X^k_j) + EI(X - X^k_j \to X^k_j)$$

And as we can see, the same analysis can be done for any pair of subsets within the whole system. We start with bipartitions that contain just one or two elements and end with a subset corresponding to the entire system. Thus, we can search for a set of bipartitions [Y, Z] within X for which $EI(Y \leftrightarrow Z)$ reaches a minimum. This bipartition indicates the information that is integrated within a system X considered as a single entity. To be comparable over bipartitions, $EI(Y \leftrightarrow Z)$ should be normalized by $EI(Y(random) \leftrightarrow Z(random))$ that substitutes bias and reflects background statistical causal relations between two random processes. Such normalized minimum of effective information for X is called *information Integration* – "Φ" [2].

In case of $EI(Y \leftrightarrow Z) = 0$ we are dealing with two causally independent subsets. If a subset Y can not be partitioned any further to a causally independent subset then we will call that unity a *functional complex* or *core*. From a practical point of view, the case $EI(Y \leftrightarrow Z) = 0$ would be trivial and very rare. Usually, dynamic objects are at least weakly coupled with many other objects that we call environment.

So, we can call two subsets causally independent when information integration between them is much less than information integration among elements within each subset. Such a constraint will lead to more fuzzy bounds which can define as core elements. Depending on the current state of the whole system, some elements can join the complex or start to be causally independent that lead to temporal dynamics for functional cores. Dynamic core will be used for the complexes with their ever-changing composition yet ongoing integration.

Another aspect of a complex information system is not just dynamic integration of information but also information differentiation and hierarchy. Complex functionality such as a brain waves patterns or stack of communication protocols on Internet are examples of balance between dynamic information integration on different levels of abstraction and differentiation from low-level protocols like IP up to high-level HTTP for the Internet. Such differentiation involves a dense network of causal interactions among informational elements. Causal density [3] is another measure that we employ with Granger causal density [4] to define new measures such as *causal power* and *recursive complexity*.

Granger causality for two information objects can be demonstrated as dynamics of state machines or finite automata. Consider two time series which enumerate dynamic states of subsets x and y of information objects by numbers X(t) and Y(t) respectively. Temporal dynamics can be described by a bivariative autoregressive model:

$$X(t) = \sum_i A * X(t - i) + \sum_i B * Y(t - i) + D(t)$$

$$Y(t) = \sum_i E * X(t - i) + \sum_i F * Y(t - i) + G(t)$$

Thus, X causes Y if coefficient E is significantly different from zero and vice versa Y causes X if B is significantly different from zero. Based of this definition, the causal density of a system can be calculated as

$$CD = C / (n * (n - 1)),$$

where C is total number of significant causal interactions that captures dynamical heterogeneity among information objects and n * (n - 1) is the total number of directed edges in a fully connected network with n objects, excluding self-connections.

Complex systems often evolve toward a hierarchical internal structure. One example is the OSI model stack of communication protocols (IP, TCP/UDP, HTTP, ...). Another example from biology is a hierarchy of molecular interaction, cellular interactions and interactions among cellular networks in living tissue.

To address such evolution, we propose a new descriptor that we call *recursive complexity*. That descriptor reflects hierarchy across different levels of description within a system. It extends the idea of Granger causality and applies renormalization semi-group transformation to the time series X(t) and Y(t). We start with time series for most frequent tokens passed among elements within the system. We call it lowest level alphabet. Then renormalization is applied and we get so called second-level time series in which each state is some combination of low-level tokens.

For example, two informational sources exchange tokens. Let's assume a time series for one of them like

```
aaa bbb ccc aaa abc bbc aaa bbb ccc ....
```

At the lowest level, we will find out that the tokens aa usually will be ended by another token a. Similarly bb and cc which will be ended by b and c respectively. But at the level of patterns like triplets, we find a more complex rule such as "if triplet aaa is followed by bbb, the next triplet will be ccc. This implies higher level of causal interactions between informational objects.

We can repeat this procedure recursively and calculate causal density until it reaches zero. When we can not find the next level then we are at the *highest causal level*.

The vector of causal densities will define the *causal power* as the vector dimension. Also, this vector will represent the recursive hierarchy of causal relations within the system.

The N level hierarchy will be represented by coded strings of the form $x = (x_1, x_2, …)$ where $x_i$-causal density at level i and i = 0, 1, …, N levels starting from the highest level (0) down to the lowest (N). The hierarchy of causal densities provides p-adic algebraic structure [5] X that we called *recursive complexity*:

$$X = x_0 + x_1 * p + \ldots + x_N * p^N \ldots$$

where $p = \Phi$ is information integration for the whole system. These are so called p-adic numbers, where the order of hierarchy $x_1, x_2, \ldots, x_N$, represent temporal order of hierarchical level which join the causal network.

Each dynamic subsystem $X^i_k$ can be described by the state vector:

$$X^i_k = x_i * p^i + \ldots + x_{i+k} * p^{i+k}$$

which accumulate three aspects of a complex informational system such as information integration, hierarchy of system internal levels and the causal density at each level.

In summary, a system can be analyzed to identify its functional complexities or Dynamic Cores – subsets of elements which can integrate information among themselves, and each complex will have an associated value of the amount of information it can integrate. That value will have temporal dynamics due to ever-changing composition yet ongoing integration.

*4. Information Geometry Unified Formalism for Analysis of Informational Dynamic System*

Dynamic informational system can be characterized by certain sets of statistical parameters. The idea of endowing the space of such parameters with metric and geometrical structure has been borrowed from parametric statistics [1]. Given the a probability distribution of $p(X|Q)$, and a sample $(x_1, x_2, \ldots)$ as a causal power vector then the objective is to estimate the Q. This may be done by maximizing the so-called log-likelihood function: $\ln L(Q) = \sum \ln p(x_i|Q)$.

Based on Frobenius theorem [6, Appendix 6], we can state that vector space "X" will have manifold as a family of integrated curves. We proposed Fisher information as a metric of geometric space for p-distributions:

$$G(ij) = -1 / N * p(x|Q) * \sum [\partial^2 \ln p(x_i|Q_i) / \partial Q_i \partial Q_j] = \partial_i \partial_j f.$$

Integral information geometry representation will give us matrix tensor:

$$g_{\mu\nu} = \int dx \, p(x|Q) \, (1/p(x|Q) \, \partial p(x|Q) / \partial Q\mu) \, (1 / p(x|Q) \, \partial p(x|Q) / \partial Q\nu))$$

Using that matrix, we can define distance between two distributions that helpful to analyze Shannon information entropy dependent on probability to have X.

To introduce the invariant functional which describes the dynamics of Fisher information metric we can define affine connection and curvature tensor:

$$\Gamma^\sigma_{\lambda\nu} = \tfrac{1}{2} g^{\sigma\nu} (\partial g_{\mu\nu} / \partial Q^\lambda + \partial g_{\lambda\nu} / \partial Q^\mu - \partial g_{\mu\lambda} / \partial Q^\nu)$$

$$R^\lambda_{\mu\nu k} = \partial \Gamma^\lambda_{\mu\nu} / \partial Q^k - \partial \Gamma^\lambda_{\mu k} / \partial Q^\nu - \Gamma^\eta_{\mu\nu} \Gamma^\lambda_{k\eta} - \Gamma^\eta_{\mu k} \Gamma^\lambda_\nu$$

As well as other geometrical tensors such as
Ricci tensor: $R^{\lambda}_{\mu\lambda k} = R_{\mu k}$
Curvature scalar is given by $R = g^{\mu k} R_{\mu k}$

Then we consider functional

$J = -1/16\pi \int \sqrt{g(Q)} R(Q) dQ$ , where $g(Q) = \det g^{\mu k}$ -is invariant for coordinate transformation

and we assume least action principal and consider small virtual fluctuation of metric that lead to Euler-Lagrange equations.

Adding to J invariant scalar term dependent on arbitrary covariant tensor

$J = -1/16\pi \int \sqrt{g(Q)} R(Q) dQ + 1/2 \int \sqrt{g(Q)} T^{\mu k} g_{\mu k} dQ$

will lead well known Einstein equation for general theory of gravitation.

$R^{\mu k}(Q) - g^{\mu k}(Q) R(Q) + 8\pi T^{\mu k}(Q) = 0$ where $T^{\mu k}(Q)$ –is constraints on the statistical system.

Another J functional with function f –gradient vector field defined on the manifold:

$J = \int (R + |\nabla f|^2) \exp(-f) dV \to \int (R + |\nabla f|^2) dm$ (well know from string theory where it describes the low energy effective action) may be considered as the gradient flow $dg_{ij}/dt = -2(R_{ij} + \nabla_i \nabla_j f)$ that is generalization of the Ricci flow $dg_{ij}/dt = -2R_{ij}$.

Another physical analogy would be recall that the partition function for the canonical ensemble at temperature $1/\beta$ is given by $Z = \int \exp(-\beta E) dw(E)$, where $w(E)$- is a "density of states" measure. Then one computes the average energy $<E> = -\partial/\partial \beta \log Z$ and the entropy $S = \beta <E> + \log Z$ for $\log Z = \int (-f + n/2) dm$

$<E> = -t*t \int (R + |\nabla f|^2 - n/(2t)) dm$
$S = -\int (t(R + |\nabla f|^2) + f - n) dm$

This statistical analogy is related to the description of the renormalization group flow and also Ricci flow. Interesting thing about Ricci flow is that it can be characterized among all other evolution equations by infinitesimal behavior of the fundamental solutions of the conjugate heat equation.

Finally, another evolutional equation that can be introduced as a candidate for delocalization is Calabi flow which, unlike the Ricci flow, is only defined on Kahler manifolds with complex coordinates $(z_i, z'_j)$:
$\partial g_{ij}/\partial t = \partial R/\partial z_i \partial z'_j$ .

A physical context for the Calabi flow in general relativity represent spherical gravitational waves in vacuum. Define informational metric as
$g_{ij} = 2(\exp(\Phi(z, z', t)))$ – exponent of information integration then two flows assume the same form

Ricci: $\partial\Phi/\partial t = \Delta\Phi$
Calabi: $\partial\Phi/\partial t = -\Delta\Delta\Phi$

That analysis lead to conclusion that any information process can be represented as a geometrical flow or visa verse any geometrical flow can be re-interpreted (isomorphic) to an informational statistical system.

5. Geometric Flow

Summarize previous chapter, we propose model hypothesis that an *architecture transition can be seen as a geometrical flow of information entropy on some informational manifold*. Results [7] demonstrates that Ricci Flow can be considered as renormalization semi-group that distribute informational curvature over the manifold but keep invariant $R = R_{min} * V^{2/3}$ where R-curvature and V-volume on information manifold. Region with strong curvature interpreted as sub-system with high information integration and recursive complexity. Thus, we proposed Ricci flow as a process of delocalization that provides distributed representation under architecture transition. Based on Perelman works [7] for the solutions to the Ricci flow ($d/dt\ g_{ij}(t) = -2R_{ij}$) the evolution equation for the scale curvature on Riemann manifold

$$d/dt\ R = \Delta R = 2|Ric|^2 = \Delta R + 2/3\ R^2 + 2|Ric^o|^2$$

implies the estimate
$R^t_{min} > -3/(2*(t+1/4))$
where the larger t-scalar parameter then the larger is the distance scale and smaller is the energy scale.
The evolution equation for the volume is
$d/dt\ V < R_{min} V$.
Take R and V asymptotic at large t, we have
$R(t)V(t)^{-2/3} \sim -3/2$
Thus, we have Ricci flow as a process of delocalization where V-growth when R-decrease that provides distributed representation under architecture transition.

*6. Topological Torsions, Pfaff Dimension Coherent structures*

To analyze global properties of informational systems – universes, we consider topological properties of informational manifold. A coherent structure such as Dynamic core is viewed as a deformable connection domain on information manifolds with certain similar topological properties. We consider topological properties of such space that stay the same under continuous transformations. The fundamental problem is how to extract topological information from information flow.
Lets' consider a system where recursive complexity X evolves in time with some velocity VX. Following the method of Cartan [8], a certain amount of topological information can be obtained by the construction of the Pfaff sequences based on the 1-form of Action, $A = ds = VX_\mu(x)dx^\mu$ a differential constructed from the unit tangent information integration velocity field VX.

We claim that emergent states are coherent topological structures:
Topological Action (energy) A
Topological Vorticity (rotation) F=dA
Topological Torsion (entropy) H=A ∧ dA
Topological Parity (Dynamic Core) K = dA ∧ dA

The Pfaff dimension of an information *universe* is the rank of largest non-zero element of the above sequence. It gives us the minimal number M of functions required to determine the topological properties of the given form in a pre-geometric variety of dimensions N. We require at least dimension 4 to accommodate complex systems with dynamic cores.

The Pfaff dimension is an invariant of a continuous deformation of the domain then it is invariant under geometrical flow. A fundamental question is: how do systems originally of Pfaff dimension 3 or less evolves into systems of Pfaff dimension4? Such evolutionary flows involve changing topology. It may be demonstrated from deRham theorem [8] and Brouwer theorem [8] that the odd dimensional set (1,3,..) may undergo topological evolution but even dimensions remain invariant. It implies that Dynamic Core coherent topological structure once established through evolution of the Pfaff dimension from 3-to -4 then will remain invariant.

*6. Comments*

With regard to proposed quantitative measures, any measured value involves the calculation of informational entropy, which requires the identification of a set of states to which probability of occurrence can be assigned. For complex systems, the identification of such a set depends on arbitrary choices because such system can be described by different variables, such as transferred bits/bites, quantity of tokens or messages passed between elements and amount states for each object presented as finite automata. The set of informational states corresponding to each variable or to any combination of variables will, in general, be different, and therefore, the corresponding values for entropy will also be different. But we are looking for causal constraints rather then for the absolute values of parameters which are really dependent on the observer in the specification of the units in which a given variables is measured.

We are looking for a transition of the system architecture that will preserve Dynamic Core complexes and make them Transient Dynamic Cores that imply the ability to effectively re-implement functionality – information integration by ever-changing composition yet ongoing integration.

*7. Physical Metaphors*

Usually, in working with dynamic systems, an investigator is trying to analyze stability or robustness of the system in terms of differential equations. The system is considered stable or robust during some period of time if some local properties of the system can be observed.

From many physical and engineering applications we can observe examples of dynamic systems that exhibit various types of behavior such as periodic, deterministic and chaotic. That classification is based of observation local characteristics of a dynamic system with assumptions that such characteristics, first, can be observed, and second, are enough to describe the system evolution. Now, we would like to develop an equivalent approach to analyze dynamics of ultra-large scale information systems--those start to emerge in information technology.

Information conservation during Liouville dynamics [9,10] which implies that total entropy of the system is preserved and Landau principle of show the way to explore the possibility of the conservation of some integral properties of informational systems. The conservation of the total amount of information can be stated as reason because the surrounding environment – "heat bath" must contain the information being erased.

Thus, we can use dynamic system terminology to reformulate our major question in physical terms as *"what kind of symmetries should be implemented in the informational universe (environment) to provide dynamic cores conservation due to dynamic transitions defined by entropy flow."*

We recall Landau principle to employ it for our informational "universe" in respect of conservation of functionality of delocalized Dynamic Cores. We will show further in the paper that for some interactions the total information is conserved and we will build the type of non-local Hamiltonians are named as Conservative Hamiltonians, similar to conservative forces. Those Hamiltonians extended with heat bath still conserve information due to Landau principle for the whole system.

Complex systems like Dynamic Cores can continuously produce and dilute information but we will postulate conservation of the "Recursive Complexity rotor"=rot (VX) where VX= $\partial X/\partial t$ is velocity vector of Recursive Complexity evolution.

We will demonstrate that localization of the Dynamic Core can be seen as information compression. And we will propose hypothesis that biological systems evolve toward "maximum of implicit functionality" that later can emerge through information compression as explicit functionality manifested by specialized organ development.

HOLOGRAPHIC REPRESENTATION FOR DYNAMIC INFORMATION OBJECTS

*1. Delocalization as Holographic representation*

Recall the physical metaphors; we can represent a physical object as a holography. The Holography Principle developed by Witten for moving object in 5D anti-De-sitter space [11] represented as 4D holograms is a good example. We will generalize that approach for Delocalization of object on information manifold.

First, we consider Dynamic Core as an assembly of causally related information objects represented by geometric features on an informational manifold. Where each point of the geometric object represents a state of an informational agent involved in formation of the Dynamic Core. Distance among points will reflect power of causal interactions and hierarchical causality among agents.

As we mentioned before, Witten holographic solution for black holes in anti-De sitter space gives us physical metaphor that can be generalized for informational manifolds. Now we would like to connect localized representation with delocalized one.

A toy example on figure 3a shows the Dynamic Core as a geometric object of causally related objects on a 3D informational manifold and holographic representation of such object as a set of concentric conics representing individual points.

*Figure 3a. Demonstrate distributed representation of a Dynamic Core on 3D informational manifold as a set of concentric conics representing individual points. Each point of the Dynamic Core is a state of an information system and the object itself is an assembly of causally related information objects.*

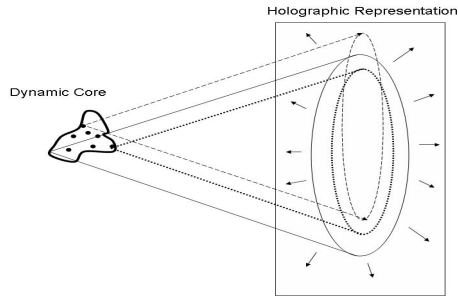

As demonstrated in figure 3, each point can be projected to a elliptic curve $R^0 \rightarrow S^1$ but we can make a more generic transformation to a sphere $R^0 \rightarrow S^2$ or even to a set of Riman spheres $R^0 \rightarrow \{S^n\}$. Transformation to 2-dimensional spheres is particularly interesting due to its association with twister geometry developed by Penrose [12]. Another transformation $R^0 \rightarrow S^1 \times S^1$ can be associated with geometry of Cremona space developed by Saniga [13].

*Figure 3b.*

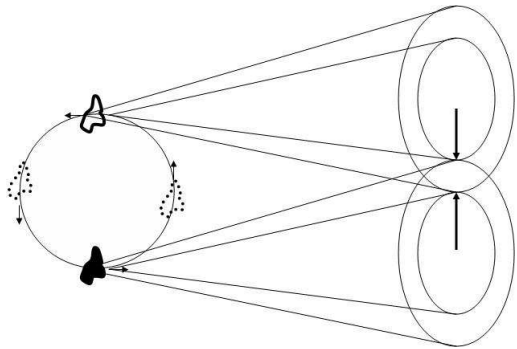

*Figure 3c.*

Now consider toy interaction between two dynamic cores that can be represented as circular movements of the geometric objects around a common center. Figures 3b and 3c illustrate such dynamics as well as the evolution of holographic representation.

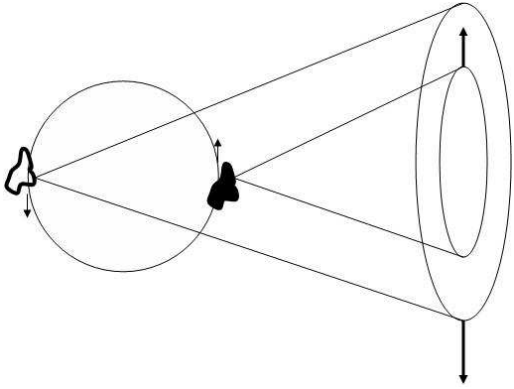

The question is what properties of the environment can produce dynamics of holographic representations which will be equivalent to circular rotation of localized Dynamic Cores.

*Figure 3d.*

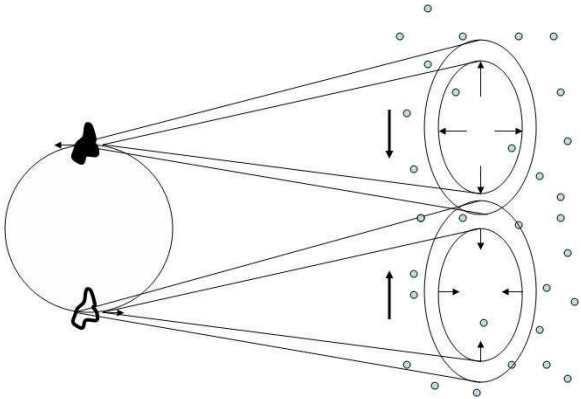

Figure 3d illustrates some dispersed none-linear media which are able to provide some sort of self-focusing effect where delocalized dynamic core is a holography wave field on the informational manifold (environment). The wave field in the inhomogeneous layer satisfies the Helmholtz equation

$$d^2/dx^2\ u(x) + k^2(x)u(x) = 0\ ;$$

where $k^2(x) = k^2(1+\epsilon(x))$ and function $\epsilon(x)$ describes the inhomogeneities of the medium. In the simplest case, we assume that $k(x) = k$, i.e., outside the layer and $\epsilon(x) = \epsilon_1(x) + i\epsilon_2(x)$ inside a layer where real component responsible for wave scattering and imaginary part for absorption. As well known from theory of dynamics and stochastic systems [14], Operate by $\epsilon_1(x)$ and $\epsilon_2(x)$ we can find conditions that perform localization of the wave. Such layered environment will be an example of the "universe" with conservative dynamic core which preserve functionality during delocalization. One way to construct such environment would be introduction of the non-linear layered media with multiplicative noise [15] that can induce stochastic resonances in the system. Non-linearity of the media will lead to self-focusing of the propagating waves and resonances will provide a mechanism to increase the wave's intensity.

$$\partial\Phi/\partial t = \partial^2\Phi/\partial x^2 + f(\Phi) + \xi(x,t)$$
$$\partial\Phi/\partial t = \partial^2\Phi/\partial x^2 + f(\Phi) + g(\Phi)\,\xi(x,t)$$

Such a self-focusing in term in information dynamics meant information concentration that requires increase of entropy in the local area of informational manifold.

On the other hand, work [15] demonstrates how a flux density can be localized within randomly layered media. Application of that formalism for informational manifold provides the method to find out conditions for localization (compression) of delocalized dynamic core.

Figure 4. Dynamic information system can be seen in local and holographic representation.

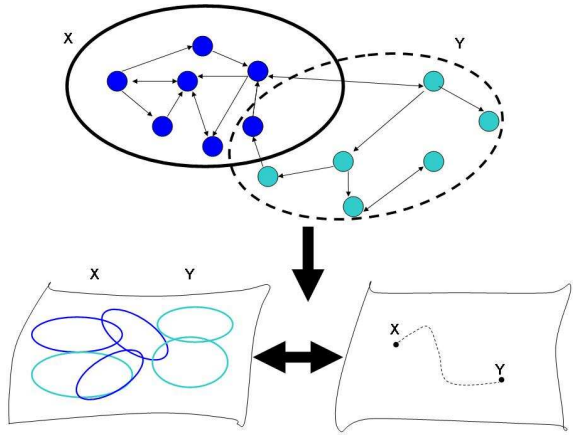

The effect of information concentration and dilution is known for non-local Hamiltonians in a finite Ising model [16]. Thus, we can conclude that environment that define informational manifold equivalent to the one for informational geometry of finite Ising model will be good candidate for holographic representations of our toy Dynamic Cores.

*2. Matrix Model for Holographic Delocalization of Dynamic Cores*

Ricci flow was proposed as an information entropy flow that conserves dynamic core functionality by providing holographic representation for the initially localized core. Also we proposed various approaches to build information *universes* which may support conservative delocalization of dynamic cores. One of that formalism for holographic representation is based of Smolin [17] matrix model with hidden variables.

Any holographic representation sub-system $X_a^b$ (figures 2, 3) can be seen as a matrix N*N –of N of informational objects, where diagonal "ii" represent holographic set of projected eclipses and internal information integration for "i" if it has internal structure. And off-diagonal element "ij" represent effective information between elements i and j.

Interactions among sub-systems can be expressed by an Action $S \sim m \int dt \, Tr\{X'^2_a + \omega^2[X_a X_b][X^a X^b]\}$ where $X_a$ is N*N matrix that can be represented X=D+Q as sum of diagonal D=diag($d_1,d_2,..$) and none-diagonal pieces. Then action can we wrote as $S \sim m \int dt \, (L^D + L^Q + L^{int})$

$L^D = m \sum D'^2_a$

$L^Q = m\{\sum Q'^2_a + \omega^2[Q_a Q_b][Q^a Q^b]\}$

Nelson's stochastic formulation of quantum theory emerges naturally as a description of statistical behavior of the eigenvalues.

$L^{int} = U(D,Q)$ - potential of interaction between diagonal and none-diagonal elements.

$L^{int} = 2m \omega^2 \sum \{-(d_i-d_j)^{2a} Q^2_b - (d_i-d_j)_a (d_i-d_j)_b Q^a Q^b - 2(d_i-d_j)\,^aQ^b [Q_a Q_b]\}$ based on this potential the classical equation of motion can be wrote how each matrix element moves in an effective potential created by the average motion of other elements.

We assume that statistical averages satisfy (Gaussians processes) relations consistent with the symmetry of the theory. This gives us

$<Q^{ij}_a Q^{kl}_b> = q^2 \delta_{ab} (\delta^{ik} \delta^{jl} + \delta^{il} \delta^{jk})$

$<(d^i-d^j)_a (d^k-d^l)_b> = r^2 \delta_{ab} (\delta^{ik} \delta^{jl} + \delta^{il} \delta^{jk})$

$<Q> = <D> = 0$

Due to gauss processes that can be formulated as diagonal elements perform Brownian motion in potential created by interaction with non-diagonal elements. From random matrix theory follow that eigenvalues undergo Brownian motion also.

That gives use Brownian movements in potential:

$<U> = m\Omega_Q^2/2 \, Q_{ija} Q^{ija} + m\Omega_d^2/2 \, (d_a^i-d_a^j)^2$ where $\Omega_Q^2 = 4(d-1)\omega^2 [(N-1) q^2 + 2 r^2]$

$\Omega_d^2 = 4(d-1)\omega^2 q^2$

The Q system is in distribution.
Variation principle for Matrix Model can be reformulated for eigenvalues.

$\lambda = d + \sum Q^{ij}_a Q^{ji}_a / (d_i-d_j)_a + \ldots$

Diffusion constant for the eigenvalues is $\nu = (\Delta d)^2/\Delta t$
Now we would like to show that neural-network can emulate quantum mechanical system at normal temperature. We are going to make one assumption that $T/(8(d-1)m \omega^2) = t/N^p$

$\hbar = m\,v$

$\Psi = \sqrt{\rho}\exp(S/\hbar)$, $\rho$ - probability density = $1/Z \exp(-H(Q)/T)$

$H(Q) = m\{\sum Q'^2_a - \omega^2[Q_a Q_b][Q^a Q^b]\}$

Lee Smolin demonstrates that variation principle in presence of Brownian motion equivalent to Schrödinger equation.

$i\hbar\, d\Psi/dt = \{-\hbar/2m\, d^2/d(\lambda)^2 + m\Omega_d^2/2 \sum (\lambda_a^i - \lambda_a^j)^2 + TN(N-1)/4 + NmC\}\,\Psi$

That analysis provides analogy between delocalized holographic representation of the dynamic core and effects of quantum theory. We can generalize such approach and analyze bundles on informational manifolds using Yang-Mills gauge theory [Appendix C]. Another example is Ising model developed on information manifold.

*3. Ising modes for Information Geometry*

As we mentioned before in our toy example (Figure 3), Ising model can be an environment reach enough to provide delocalized representation for dynamic cores. We explore that hypothesis in our work.

We introduce coordinates for our informational manifold:
$X = X_m\Phi^m + X_{m-1}\Phi^{m-1} + X_{m-2}\Phi^{m-2} + \ldots$, where $X_m$ – is causal density of the m-th level, $\Phi$- information integration and m-is number of hierarchical levels.

The geometric structure in which informational manifold is endowed leads to certain local invariants, one of the most important being the Ricci scalar curvature R.
$R \sim \zeta^d$, where $\zeta$-is the correlation length, which is two-point function, and d- denotes the number of spatial dimensions of the model. The curvature also receives contribution from higher order correlations.

The scaling behavior of the curvature in the vicinity of the critical point provides a satisfying picture of how certain universal features of the near-critical regime are encoded in the Fisher-Rao geometry of the informational manifold.

We will call N*-critical size, above which the system is to a reasonable approximation 'thermodynamic'. When the size drop below N* the system often behaves in qualitatively different way. R is strictly positive in the thermodynamic regime and negative at none-thermodynamic regime.

We can start with the state of the system immersed in a large heat bath with fixed temperature T in thermal equilibrium. The Gibbs measure can be mapped to Hilbert space
$P(x|\theta) = \exp(-\sum_{i=1:r} \theta_i H^i(x) - \ln Z(\theta)) = \psi_\theta(x)$ where $\{\theta_i\}$-represent r-dim sub-space S in the Hilbert space. Fisher –Rao metric can be introduced on the maximum entropy manifold as :
$g_{ij} = 4\int \partial_i \psi_\theta(x)\, \partial_j \psi_\theta(x)\, dx = \partial_i \partial_j \ln Z(\theta) = \partial_i \partial_j S(entropy)$, $\partial_i = \partial/\partial\theta^i$. Entropy $S(p|q) = \int p \ln(p/q) = S(p(\theta)|p(\theta+d\theta)) = 1/2\, g_{ij}\, d\theta^i d\theta^j + \ldots$Teylor series
And manifold itself interpreted as a maximum entropy surface and consequently specific geodesics will correspond to equations of state for the system.

If we take simple (Ising chain) source-to-source (H1) and source to net (H2) interaction then r=2 and
$-\beta H = \beta \sum s_i s_j + h \sum s_i$, $\beta = 1/kT$, $s=\{+1,-1\}$, h-network "field"
Scalar curvature $R = -1/(2\det(g)) * \det\{$

$$\begin{matrix} \partial_1^2 \ln Z & \partial_1 \partial_2 \ln Z & \partial_2^2 \ln Z \\ \partial_1^3 \ln Z & \partial_1^2 \partial_2 \ln Z & \partial_1 \partial_1^2 \ln Z \\ \partial_1^2 \partial_2 \ln Z & \partial_1 \partial_1^2 \ln Z & \partial_2^3 \ln Z \end{matrix}$$
$\}$
and act as an indicator of finite size effects. Differentiate $N^{-1}$ in Z we have
$g_{ij} = 1/N \, \partial_i \partial_j \{N \beta + \ln [(\cosh h + \eta)^N + (\cosh h - \eta)^N]\}$, where $\eta = (\sinh^2 h + \exp(-4\beta))^{1/2}$

Thus the size of dynamic core depends on temperature as $T \sim (1/\text{size})$ or $T \sim \text{size}$. Through Ricci flow $R<0$ (localized DC) can transit to $R>0$ delocalized implementation.
if $N \to \infty$ then we have thermodynamic curve $R = 1 + \eta^{-1} \cosh h > 0$.

Figure 5. Process of Holographic representation can be seen as Geometric flow that smooth and delocalize curvature. Region with strong curvature interpreted as sub-system with high information integration and recursive complexity

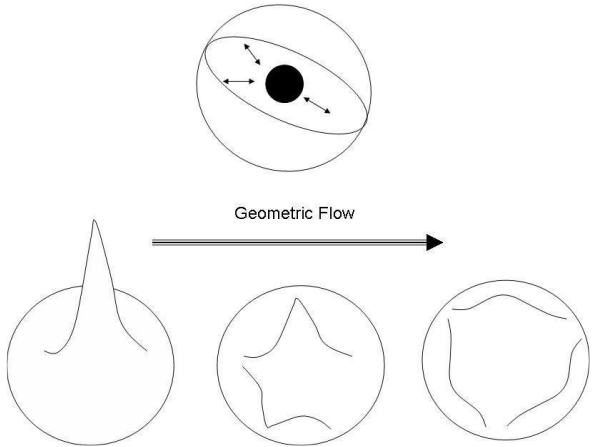

Now, lets consider just two sources with interaction described by a Hamiltonian:
$H = \sum C_i * \sigma_i * \tau_i$.
The prescribed state is denoted by the density matrix $\rho = |\psi\rangle\langle\psi|$, where $|\psi\rangle = \cos(\theta/2)|0\rangle + \exp(i\varphi)\sin(\theta/2)|1\rangle$ is superposition of active $|1\rangle$ and $|0\rangle$ none-active state. As we can see an initial state of the system of two sources is $|\rho(0)|_{4 \times 4}$. The information encoded in the system at time t is characterized by the fidelity $F_i(t) = \langle \psi | \rho^i(t) | \psi \rangle$, where $\rho^i(t) = \text{Tr } \rho(t)$-except "i".
Define two quantities:
$CF_i(t) = \cos^2(\theta/2) \rho^i_{00}(t) + \sin^2(\theta/2) \rho^i_{11}(t)$
$QF_i(t) = \text{Re}[\exp(-i\varphi) * \sin(\theta) \rho^i_{10}(t)]$
And rewrite Fidelity as $F_i(t) = CF_i(t) + QF_i(t)$, where $\rho(t) = \exp(-iHt) * \rho(0) * \exp(iHt)$.
By straightforward calculations, we found that if $C_i = C_j$ (uniformity of information integration among all sources) for any i and j then $\sum CF_i(t)$ and $\sum QF_i(t)$ are both invariable. That mean total $F(t) = \sum F_i(t)$ is invariable too.

Paper [16] prove information conservation theorem for quantum systems but using same mathematical formalism that we employ for our Dynamic core.

Due to the interactions between parts of the system, the states of these parts changes with time, and the information is expanded between them. But the total Fidelity-information (information integration) is conserved. This is something like Energy Conservation Law for information systems. This law leads to interesting phenomena when information can partially concentrated spontaneously in one part of the whole system due to oscillation part of $F_i(t)$.

Such concentration and following dilution in large-scale system with thermal bath can be seen as localization and de-localization of the Dynamic Core.

The theorem [16] also gives us a mechanism to construct informational systems (universes) with conservative dynamic core. Such universes should have "conservative Hamiltonian" that will preserve total information integration. Hamiltonian should satisfy condition $[H, \sum_{i=1}^{n} C_i] = 0$, where $C_i = 1/(2^{n-1}) I_1 * I_2 * \ldots \rho^0_i * I_{i+1} * I_{n+2} * \ldots I_n$, with $\rho^0_i$ denotes the reduced density matrix of the i-th source and $I_i$ denotes reduced density matrix of other sources. This is easily can be proven by showing that $\partial F(t)/\partial t = 0$ is equal to $[H, \sum_{i=1}^{n} C_i] = 0$.

*5. Experimental Simulation*

We performed simulation study to address the problem of optimal topology for the network of the informational objects to achieve highest information integration and differentiation.

State diagram was considered for N-informational objects which can communicate by message exchange. Our simulations demonstrated that free-scale network with gamma 2.1-2.4 where each node generate gauss distribution of tokens achieve highest information integration and provide maximum stability for Ricci entropy flow. Also, topological dimension constrain were found for informational manifolds. We found that manifolds with Pfaff dimensions 3-12 provide Dynamic Cores conserved under delocalization performed as geometric flow.

**CONCLUSION**

We propose the use of an information geometry approach that seems natural to investigate non-local properties of the system Dynamic Core or environment. That informational formalism seems more relevant for analysis of the system causal power, information integration and hierarchy. Information geometry focused on informational characteristics of dynamic systems such as information integration, hierarchy of system internal levels and the causal density at each level. Entropy dynamics can represent actions of finite state automata and their message exchanges while the energy will reflect the number of states that system might have.

Delocalization of DC possible if transition satisfied holographic constrain: delocalization N dimensional dynamic core conserve functional properties only if exist N+1 compact representation of DC in N+1 dim informational manifold that has holographic reduced representation in N-dim hyperspace equivalent to original DC manifold.

Several models was proposed for analysis of open information systems such as computer networks but the mathematical formalism is general enough and can be applicable for analysis of any dynamical system. We can speculate about *delocalization* in biological neural networks that might explain the recovery of some functions lost during stroke or cancer. One interesting example would be the story in which tumor was removed from brain area related to speech. The patient ability to speak was lost due to destruction of the area but then ability to speak was recovered in several months. One proposed explanation might be that other areas of the brain recovered the function through its delocalization.

Topological dimension constrain 3 to 12 were found for informational manifolds with conserved Dynamic Cores (where delocalized N-dim DC is a self-assembling topological knot in dim N+1) imply some technical insight for area of Artificial Intelligent and robotics. The system should have from 3 to 12 perceptual modalities like "viewing", "hearing", and "smelling", and so on to be stable but flexible enough to perform self-organized dynamic regimes and continuously learn new type of tasks. Such dimensional constrain is an optimum that allow an informational dynamic system survive in changing environment.

The good physical analogy would be the process of percolation where fluid (information) is flow through the porous media. If there is too few porous then fluid flow can not go through but if there is too many porous then the fluid just go through the media. Otherwise, if number of porous in some region of values (depending on fluid and media) then we can observe very complex dynamic patterns.

It is also might lead to another biological speculation about number of sensor modalities allowed in variety of species. It is interesting fact from biology that we probably do not have species with less then three and more then 7 sensor organs. Dimension constrain gave some formal explanation of that fact. If amount of perceptual modalities less then 3 then system does not have enough internal complexity and none-linearity to build reach and self-conservative internal representation of environment and it-self interactions. That led to poor flexibility, none-robustness and poor ability to survive (Dynamic Cores does not conserved) during environment transitions. Otherwise if it has too many perceptual modalities it makes the system too complex (Dynamic Cores also does not conserve) and non-robust due to slightest fluctuations in information transfer.

Finally, we would like to mentioned possibility of link between resurrection of functional properties in delocalized dynamic cores and processes of live in which biological functionalities emerge at different hierarchical levels from local properties of tiny biological molecules.

*Appendix A. Information entropy "H"*

Crucial part of that formalism is information entropy. Information theory provides different approximation models to calculate that function. For statistically independent tokens or signals those have equal probability it provide Zero-Order Model:

*$H = log_2 N$ – where N – number of states*

If token are statistically independent but might have different probability to occur then entropy described by First-Order Model:

$H = -\sum_i p(x_i) \log_2 p(x_i)$ –where $p(x_i)$ is probability to detect signal $x_i$

One example would be probability to detect token/signal with pre-defined frequencies in stochastic system with non-uniformly distributed spectrum. That non-uniformity will lead to different detection probability for signals of various frequencies.

In case of multi channel communication in distributed enterprise system we can have situation where different channels can be correlated. The tokens will be described by $x_i$ - n-dimensional vector where n is number of communicational channels.

There is more complex situation for systems with memory where source-informational object remember its previous state (very typical for finite-automata and state-machines). It is more likely that the current token will depend on the previous one but it is conditionally independent of all other signals. This called Second-Order Model any system that goes under Markov processes will have such entropy:

$H = -\sum_i p(x_i) \sum_j p(x_j | x_i) \log_2 p(x_j | x_i)$ –where $p(x_j | x_i)$ is conditional probability. That is probability to detect signal $x_j$ after signal $x_i$
$p(x_j | x_i) = p(x_j, x_i) / p(x_i)$

If system has long memory and a token will strongly depend of two or more previous states then Third-Order or General Models should be used:

$H = -\sum_i p(x_i) \sum_j p(x_j | x_i) \sum_k p(x_k | x_j, x_i) \log_2 p\, p(x_k | x_j, x_i)$ –where $p(x_k | x_j, x_i)$ is conditional probability to detect signal $x_i$ after signal $x_j$ coming after $x_k$

*Appendix B. Frobenius Theorem for Informational Manifolds.*

A smooth vector field X on a p-adic informational manifold M of causal densities can be integrated to define a one-parameter family of curves. The integrability follows because the equation defining the curve is a first-order ordinary differential equation, and thus its integrability is guaranteed by the Picard-Lindelof theorem. Indeed, vector field will be defined to be the derivatives of a collection of smooth curves. It leads to rise of regular foliation.

Based of version of Frobenius theorem that if Li-brackets of all pair-wise combinations of m- vector fileds {X} is linear combinations of {X} then integral curves of those fields define foliation as a sub-manifolds with dimensions k<=m. Each sub-manifold called leave of foliation and vector fields define bundle.

*Appendix C. Yang-Mills gauge theory and spontaneous symmetry breaking for informational Fisher lagrangians*

Tangent vector "A" can be defined for each point X of manifold M(X) as

$$A_\mu \sim \partial \ln(p(x))/\partial x_\mu$$

Where Li brackets $[A_\mu A_\nu] \sim A_k$, give use the way to find transformations which will keep DC conservative.

First we can employ approach developed in gauge theories that are usually discussed in the language of differential geometry that make it plausible to apply for informational geometry. Mathematically, a *gauge* is just a choice of a (local) section of some principle bundle. A gauge transformation is just a transformation between two such sections.

Note that although gauge theory is mainly studied by physics, the idea of a connection is not essential or central to gauge theory in general. We can define a gauge connection on the principal bundle. If we choose a local basis of sections then we can represent covariant derivative by the connection form $A_\mu$, a Lie-algebra valued 1 –form which is called the gauge potential in physics. From this connection form we can construct the curvature form *F*, a Lia-algebra valued 2-form which is an intrinsic quantity, by

$$F_{\mu\nu} = \partial_\mu A_\nu - \partial_\nu A_\mu - ig[A_\mu\ A_\nu]$$

$[D_\mu\ D_\nu] = -ig\ F_{\mu\nu}$, where $D_\mu = \partial_\mu - igA_\mu$ ; $A_\mu = A_\mu^a\ t^a$ and t –is generator of infinitesimal transformation

We can write invariant Lagrangian: $1/4\ F_{\mu\nu}^a F_{\mu\nu}^a \Leftrightarrow -1/2\ Sp(F_{\mu\nu} F_{\mu\nu})$ that is invariant under transformation of coordinates.

Another one is
$$L = (\partial_\mu \Phi\ \partial_\mu \Phi) - m/2(\Phi^*\Phi)$$

that is symmetric under transformation $\Phi \to \Phi * \exp(i\alpha)$
Let's consider potential that break symmetry: $V(\Phi) = \lambda/4(\Phi^2 - \eta^2)^2$ ;

$L = (\partial_\mu \Phi\ \partial_\mu \Phi) - \lambda/2(\Phi^*\Phi - \eta^2)$, define $\Phi = \eta + (\chi + i\psi)/\sqrt{2}$ then will become $V(\Phi) = \underline{\lambda\eta^2\chi^2} + \lambda\eta\chi^3/\sqrt{2} + \lambda\eta\chi\psi^2/\sqrt{2} + \lambda\chi^4/8 + \lambda\psi^4/8 + \lambda\chi^2\psi^2/4$

Each infinitesimal transformation that break symmetry represents scalar "particle" - $\psi$ and direction where $V(\Phi)$ conserved. At the other hand if we add tangent vector then lagrangian

$L = (D_\mu \Phi\ D_\mu \Phi) - \lambda/2(\Phi^*\Phi - \eta^2) - 1/4\ F_{\mu\nu} F_{\mu\nu}$ ; if we define $\Phi = \eta + \chi/\sqrt{2}$ then
$L = 1/2(\partial_\mu \chi\ \partial_\mu \chi) - \underline{\lambda\eta^2\chi^2} - \lambda\eta\chi^3/\sqrt{2} - \lambda\chi^4/4 - 1/4\ F_{\mu\nu} F_{\mu\nu} + \underline{e^2\eta^2\ A_\mu A_\mu} + \sqrt{2}\ e^2\eta\ \chi A_\mu A_\mu + e^2\chi^2/2\ A_\mu A_\mu$

Such lagrangian describe spontaneous symmetry breaking with informational boson $A_\mu$ of mass $\sqrt{2}e\eta$ and neutral scalar boson $\chi$ (Higgs boson) with mass $\sqrt{2}\ \lambda\eta$ which interact with each other.

Interesting effect, such informational "particle" provides classification of various informational processes which can emerge in complex systems. Emergence of informational "mass" leads to limited depth of penetration for information integration to environment.

Spontaneous supper symmetry-breaking for string theory action can provided mechanism that limit conservative Dynamic Cores and their holographic representations to the sub-set of informational manifolds with topological space dimension from 3 to 12.

We find out that dim=3-12 condition for minimal tangent space will lead to manifolds with such constraints that entropy flow (geometric Riccy flow) perform holographic representation of dynamic cores. We will call such systems as "immortal dynamic cores".